\documentclass[conference]{IEEEtran}

\usepackage[a4paper, left=0.75in,right=0.75in,top=1.444in,bottom=0.75in]{geometry}

\usepackage{setspace}
\usepackage{balance}
\usepackage{caption}
\usepackage{graphicx}
\usepackage{xcolor,colortbl}

\begin{document}
%\huge 
\title{\vspace{0.25in} \huge Towards Efficient RRAM-based Quantized Neural Networks Hardware: State-of-the-art
and Open Issues. \vspace{-0.15in}}

\author{
\IEEEauthorblockN{ O. Krestinskaya, L. Zhang and K.N. Salama}
\IEEEauthorblockA{King Abdullah University of Science and Technology, Saudi Arabia} \vspace{-1.5cm}}

\maketitle

\begin{abstract}
\setstretch{0.9}
The increasing amount of data processed on edge and demand for reducing the energy consumption for large neural network architectures
have initiated the transition from traditional von Neumann architectures towards in-memory computing paradigms.  Quantization is one of the methods to reduce power and computation requirements for neural networks by limiting bit precision. Resistive Random Access Memory (RRAM) devices are great candidates for Quantized Neural Networks (QNN) implementations.  As the number of possible conductive states in RRAMs is limited, a certain level of quantization is always considered when designing RRAM-based neural networks. In this work, we provide a comprehensive analysis of state-of-the-art RRAM-based QNN implementations, showing where RRAMs stand in terms of satisfying the criteria of efficient QNN hardware.  We cover hardware and device challenges related to QNNs and show the main unsolved issues and possible future research directions.
\end{abstract}


\IEEEpeerreviewmaketitle

\vspace{-0.2cm}

\section{Introduction}
\vspace{-0.15cm}
%\IEEEPARstart{T}{his} 
\setstretch{0.89}
The exponentially growing amount of processed data, IoT applications, and on-edge processing imposed the requirements to develop energy-efficient low-power neuromorphic hardware.
RRAM devices are one of the most promising technologies
for neuromorphic circuits, neuro-inspired architectures, and neural network accelerators.
RRAMs allow the efficient crossbar-based implementation of neural network weights and
multiply-and-accumulate (MAC) operations leading to higher energy efficiency, smaller on-chip
area, higher processing speeds. This enables the transition from traditional von Neumann
architectures towards in-memory computing paradigms. The off-chip memory bottleneck in traditional architectures motivated the closer integration of memory and processing units to avoid frequent data transfer between the off-chip memory and processor \cite{chakraborty2020resistive}. 

 %Also, instead of using time-multiplexed architectures, where the data from the off-chip memory is transferred to the single main memory for processing (reprogramming the main memory), the spatial distributed architectures, where the memory crossbars are located across multiple cores forming a network-on-chip and having corresponding processing units, are more useful for RRAM-based implementation.   

%MAYBE FIGURE HERE!!!!!!
%The off-chip memory bottleneck in traditional architectures (Fig. \ref{f1} (a)) motivated the closer integration of memory and processing units \cite{chakraborty2020resistive} to avoid frequent data transfer between the off-chip memory and processor, as shown in Fig. \ref{f1} (b). Also, instead of using time-multiplexed architectures, where the data from the off-chip memory is transferred to the single main memory for processing (reprogramming the main memory), the spatial distributed architectures, where the memory crossbars are located across multiple cores forming a network-on-chip and having corresponding processing units, are more useful for RRAM-based implementation.  

Quantization in hardware saves memory space, reduces data movement and latency for arithmetic operations \cite{huang2021mixed}. 
Full-precision computations and complex arithmetics support for large neural networks in low-power hardware are challenging. 
Therefore, QNN is an efficient method for low-power AI applications on edge. The currently available RRAM devices still have limited precision. Thus, quantizing the computation around RRAM crossbars and implementing RRAM-based QNNs is 
a reasonable approach.
Compared to traditional in-memory computing, RRAM provides possibilities for
multi-level storage, scalability, and high computational density. For example, QNNs implemented with SRAM devices have a cell size of $124F^2$ ($F$ - technology feature size), comparing to RRAM with the size of $<10F^2$ (e.g. $4F^2$ for RRAM cells and $12F^2$ for 1T1R cells \cite{chakraborty2020resistive}).

In this work, we analyze state-of-the-art RRAM-based QNN
implementations, including performance, energy efficiency, bit-width of network weights and computations, and training methods.
We cover hardware and device challenges of QNNs, such as lack of variable bit
precision hardware and bit-width reconfigurability, training and learning limitations, demand for large-scale integration with traditional architectures, and existing
RRAM devices’ limitations. 
We also show the main unsolved issues and summarise possible
future research directions.

%Considering applications deployed on edge or embedded systems have high privacy requirements. Thus, there is a high demand on edge training. 

%High precision and complex arithmetics support -> issue -> QNN

%*FPGAs

%-- often BNNs are implemented with SRAMs but the cell size is $124F^2$, where $F$ is technology feature size, comparing to RRAM with the size of $<10F^2$ (e.g. $4F^2$ for RRAM cells and $12F^2$ for 1T1R cells \cite{chakraborty2020resistive}).

%The cost of data transfer and limiting energy efficiency of data buffers

%Architecture: 
%--reusing crossbar as in von Neumann architectures is expensive -> spartial architectures (here or in the beginning?)
%--distributed architecture, e.g. network-on-chip

%\cite{roy2020memory}: SOME DATA for reference: 
%--RRAM can overcome off-chip memory access bottleneck (However, expensive memory writes) THerefore, crossbar reuse across the layers is not an efficient approach for NVMs. A spatial architectures with multiple cores and crossbar corresponding to each core is more efficient.

%BNN:
%where accumulation operation can be transferred to the counting '1's in XNOR outputs \cite{sun2018xnor}
%Power-of-two scheme is based on the weight distribution 
%Fixed point - less overhead comparing to float
%Connection to RRAM devices

%Other NVM devices: PCM, Spintronic devices, FeFETs. Benefits of RRAM over other NVM devices: multi-level storage, scalalbility, high density, high Ron/Roff ratio \cite{chakraborty2020resistive}.

\vspace{-0.15cm}
\section{Background}
\vspace{-0.15cm}

\subsubsection{\textbf{QNN basics}}

%Quantization is a technique to compress the hardware requirements, especially for on-edge and IoT applications.
There are two approaches to quantize a neural network: 
post-training quantization (PTQ) of a full precision model
and quantization-aware training (QAT). QAT usually achieves higher accuracy than PTQ.  The application of the traditional stochastic gradient descent (SGD) training algorithm on QNNs is challenging due to the stair-like nature of the quantization function. Therefore, the straight-through estimator (STE) technique is used for training, which still requires full precision for update computation.

\begin{figure}[!t]
    \centering
    \includegraphics[width=79mm]{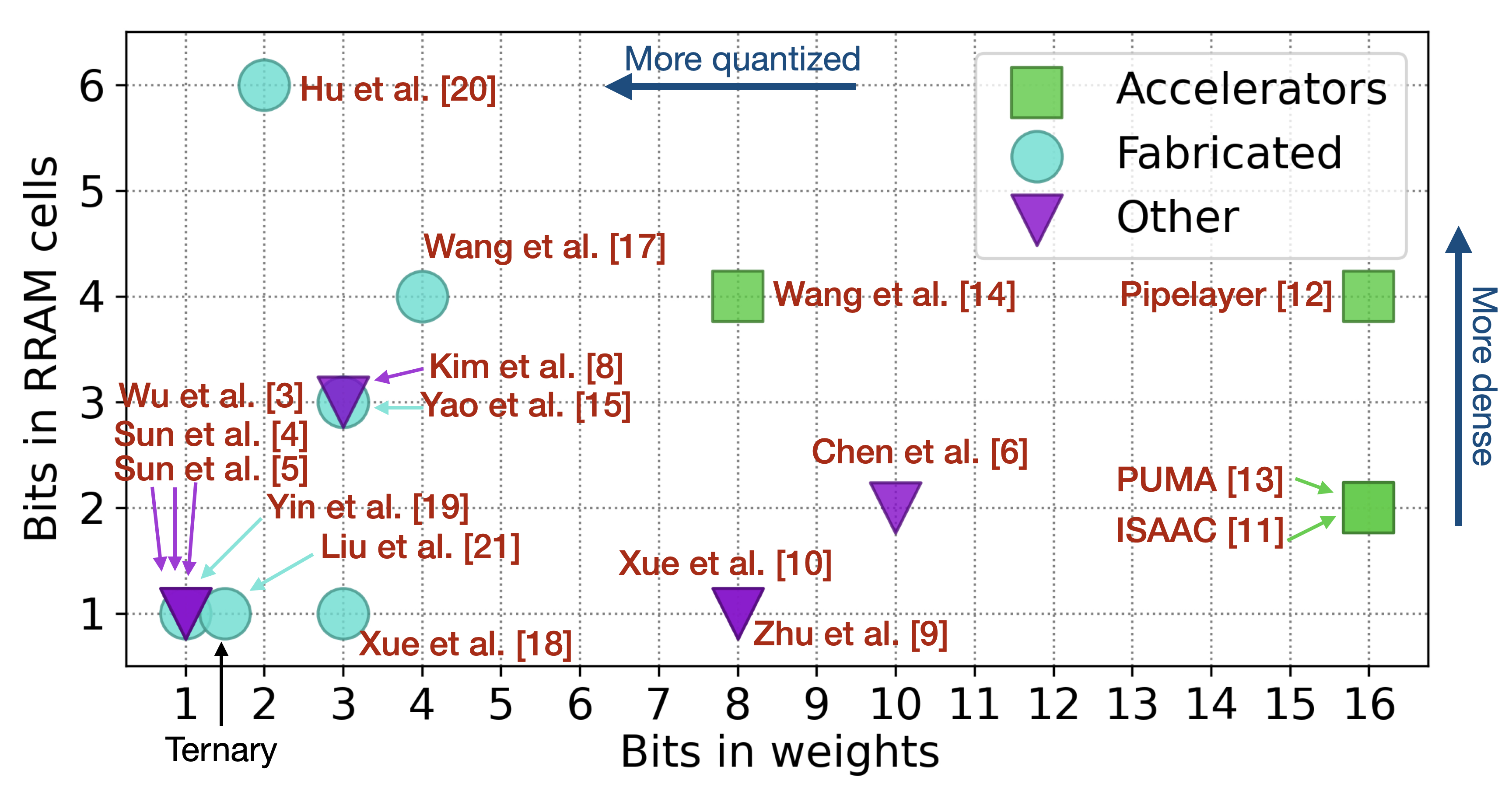}
    \caption{\footnotesize Bits per RRAM cell and weights quantization in state-of-the-art RRAM-based QNN architectures.}
    \label{f1}
    \vspace{-0.5cm}
\end{figure}

\subsubsection{\textbf{RRAM basics and related challenges}}
%RRAM is a two-terminal resistive switching device with a metal-insulator-metal structure and multiple conductive states, where the 
%switching mechanism is either filamentary, driven by oxygen migration, or non-filamentary, controlled by the barrier \cite{chakraborty2020resistive}.

RRAMs allow to achieve high density of the architecture and energy-efficient computation but are prone to non-idealities, such as device-to-device (D2D) and cycle-to-cycle (C2C) variations, low endurance, a limited number of conductive states, conductance drift with time, and device faults.
Low-bit RRAM devices have fewer variations \cite{huang2021mixed}, but lead to lower crossbar density and area efficiency.
Currently available devices have $2^6$ conductive states, $R_{on}/R_{off}$ ratio of $10$-$10^3$ with $R_{on}$=$10^3$-$10^4$. The set/reset time and write energy reaches as low as 50$ns$ and 2$nJ$ \cite{chakraborty2020resistive}, the endurance reaches up to $10^6$ cycles, and retention time up to $10^8$s. 
RAM-based 1T1R or 1T1S crossbars for MAC operation can have linear non-idealities, such as wire resistance and source/sink resistances contributing to the IR drop, and non-linear non-idealities, such as exponential dependence of current on voltage, non-linear behavior of selectors devices/ transistors and RRAMs, and non-linear asymmetric dynamics in switching behavior leading to errors \cite{chakraborty2020resistive}. The main issue of the RRAM-based crossbars is high write latency and write cost.

\begin{table}[!t]
%\centering 
\scriptsize \setlength{\tabcolsep}{0.1pt}
\renewcommand{\arraystretch}{0.55}
\caption{RRAM QNN architectures}
\begin{tabular}{|ccccccc|}
\hline
\multicolumn{1}{|c|}{\textbf{Work}}              & \multicolumn{1}{c|}{\textbf{\begin{tabular}[c]{@{}c@{}}CMOS,\\ RRAM\end{tabular}}} & \multicolumn{1}{c|}{\textbf{Tiles}} & \multicolumn{1}{c|}{\textbf{\begin{tabular}[c]{@{}c@{}}RRAM \\ bits\end{tabular}}} & \multicolumn{1}{c|}{\textbf{\begin{tabular}[c]{@{}c@{}}Bits \\ W,I,O$^{*1}$\end{tabular}}} & \multicolumn{1}{c|}{\textbf{Network}}                                                    & \textbf{\begin{tabular}[c]{@{}c@{}}Training,\\ bits\end{tabular}} \\ \hline
\multicolumn{7}{|c|}{\cellcolor{green!10} \textbf{Binary Neural Networks}}                                                                                                                                                                                                                                                                                                                                                                                                                                                                  \\ \hline
\multicolumn{1}{|c|}{\cite{9265062}}             & \multicolumn{1}{c|}{Ti/HfO2}                                                       & \multicolumn{1}{c|}{-}              & \multicolumn{1}{c|}{1}                                                             & \multicolumn{1}{c|}{1b, 1b, 1b}                                                      & \multicolumn{1}{c|}{4 FC}                                                                & \begin{tabular}[c]{@{}c@{}}QAT, 32b\end{tabular}                \\ \hline
\multicolumn{1}{|c|}{\cite{sun2018xnor}}         & \multicolumn{1}{c|}{45nm}                                                          & \multicolumn{1}{c|}{128x128}        & \multicolumn{1}{c|}{1}                                                             & \multicolumn{1}{c|}{1b, 1b, 1b}                                                      & \multicolumn{1}{c|}{\begin{tabular}[c]{@{}c@{}}6 Conv, 3 FC\end{tabular}}             & \begin{tabular}[c]{@{}c@{}}PTQ,  32b\end{tabular}               \\ \hline
\multicolumn{1}{|c|}{\cite{sun2018fully}}        & \multicolumn{1}{c|}{65nm}                                                          & \multicolumn{1}{c|}{-}              & \multicolumn{1}{c|}{1}                                                             & \multicolumn{1}{c|}{1b, -, -}                                                        & \multicolumn{1}{c|}{\begin{tabular}[c]{@{}c@{}}4 Conv,  3FC\end{tabular}}              & \begin{tabular}[c]{@{}c@{}}QAT,  6b\end{tabular}                \\ \hline
\multicolumn{7}{|c|}{\cellcolor{green!10}  \textbf{Higher precision networks (mostly fixed-point arithmetics)}}                                                                                                                                                                                                                                                                                                                                                                                                                              \\ \hline
\multicolumn{1}{|c|}{\cite{chen2018high}}        & \multicolumn{1}{c|}{65nm}                                                          & \multicolumn{1}{c|}{128x128}        & \multicolumn{1}{c|}{2}                                                             & \multicolumn{1}{c|}{up to 10b, 8b (s-1b), 5b}                                        & \multicolumn{1}{c|}{Lenet}                                                               & -                                                                 \\ \hline
\multicolumn{1}{|c|}{\cite{cai2019low}}          & \multicolumn{1}{c|}{45nm}                                                          & \multicolumn{1}{c|}{10-256$^{*2}$}    & \multicolumn{1}{c|}{1-8b$^{*2}$}                                                    & \multicolumn{1}{c|}{1-8b, 1-8b, 1-8b}                                                & \multicolumn{1}{c|}{\begin{tabular}[c]{@{}c@{}}Lenet, ResNet, \\ VGG-16\end{tabular}} & \begin{tabular}[c]{@{}c@{}}QAT, 32b\end{tabular}               \\ \hline
\multicolumn{1}{|c|}{\cite{kim20213}}            & \multicolumn{1}{c|}{-}                                                             & \multicolumn{1}{c|}{32x32}          & \multicolumn{1}{c|}{3b}                                                            & \multicolumn{1}{c|}{3b, -, -}                                                        & \multicolumn{1}{c|}{2 FC}                                                                & \begin{tabular}[c]{@{}c@{}}PTQ,  32b\end{tabular}               \\ \hline
\multicolumn{1}{|c|}{\cite{zhu2019configurable}} & \multicolumn{1}{c|}{45nm}                                                          & \multicolumn{1}{c|}{256x256}        & \multicolumn{1}{c|}{1b}                                                            & \multicolumn{1}{c|}{up to 8b, - (s-2b), 6b}                                                & \multicolumn{1}{c|}{\begin{tabular}[c]{@{}c@{}}Lenet,  ResNet, \\ VGG-16\end{tabular}} & \begin{tabular}[c]{@{}c@{}}PTQ,  32b\end{tabular}               \\ \hline
\multicolumn{1}{|c|}{\cite{xue202015}}           & \multicolumn{1}{c|}{22nm}                                                          & \multicolumn{1}{c|}{512x512}        & \multicolumn{1}{c|}{1b}                                                            & \multicolumn{1}{c|}{2-4b, - (s-1-2b), 6-11b}                                         & \multicolumn{1}{c|}{ResNet-20}                                                           & -                                                                 \\ 
\hline
\multicolumn{7}{|c|}{\textbf{\cellcolor{green!10}  Accelerators (fixed point arithmetics)}}                                                                                                                                                                                                                                                                                                                                                                                                                                                  \\ \hline
\multicolumn{1}{|c|}{\cite{shafiee2016isaac}}    & \multicolumn{1}{c|}{\begin{tabular}[c]{@{}c@{}}32nm,\\ TiO/HfO\end{tabular}}       & \multicolumn{1}{c|}{128x128}        & \multicolumn{1}{c|}{2b}                                                            & \multicolumn{1}{c|}{16b, 16b (s-1b), 16b}                                            & \multicolumn{1}{c|}{VGG}                                                                 & PTQ                                                               \\ \hline
\multicolumn{1}{|c|}{\cite{song2017pipelayer}}   & \multicolumn{1}{c|}{TiO$_{2-x}$}                                                   & \multicolumn{1}{c|}{128x128}        & \multicolumn{1}{c|}{4b}                                                            & \multicolumn{1}{c|}{16b, 16b (s-1b), 16b}                                            & \multicolumn{1}{c|}{\begin{tabular}[c]{@{}c@{}}AlexNet, VGG\end{tabular}}              & on-chip                                                           \\ \hline
\multicolumn{1}{|c|}{\cite{ankit2019puma}}       & \multicolumn{1}{c|}{32nm}                                                          & \multicolumn{1}{c|}{128x128}        & \multicolumn{1}{c|}{2b}                                                            & \multicolumn{1}{c|}{16b, 16b (s-1b), 16b}                                            & \multicolumn{1}{c|}{\begin{tabular}[c]{@{}c@{}}VGG,  LSTM\end{tabular}}                & \begin{tabular}[c]{@{}c@{}}PTQ, 32b\end{tabular}                \\ \hline
\multicolumn{1}{|c|}{\cite{wang2019deep}}        & \multicolumn{1}{c|}{65nm}                                                          & \multicolumn{1}{c|}{256x64}         & \multicolumn{1}{c|}{4b}                                                            & \multicolumn{1}{c|}{8b, 8b (s), 8b}                                                  & \multicolumn{1}{c|}{\begin{tabular}[c]{@{}c@{}}VGG-16, \\ MobileNet\end{tabular}}        & \begin{tabular}[c]{@{}c@{}}PTQ, 32b\end{tabular}                                                                                                                                                                                                                      \\
\hline

\multicolumn{7}{|c|}{\cellcolor{green!10}  \textbf{Fabricated RRAM crossbar-based architectures}}                                                                                                                                                                                                                                                                                                                                                                                                                                                                \\ \hline
\multicolumn{1}{|c|}{\cite{yao2020fully}}        & \multicolumn{1}{c|}{\begin{tabular}[c]{@{}c@{}}130nm, \\ TaOx/HfOx\end{tabular}}   & \multicolumn{1}{c|}{126x16}         & \multicolumn{1}{c|}{3b}                                                            & \multicolumn{1}{c|}{3b, 8b (s-1b), 8b}                                               & \multicolumn{1}{c|}{\begin{tabular}[c]{@{}c@{}}2 Conv, \\ 1 FC\end{tabular}}             & hybrid$^{*4}$                                                       \\ \hline
\multicolumn{1}{|c|}{\cite{li2018efficient}}     & \multicolumn{1}{c|}{HfO2}                                                          & \multicolumn{1}{c|}{128x64}         & \multicolumn{1}{c|}{-}                                                             & \multicolumn{1}{c|}{-,-,-}                                                           & \multicolumn{1}{c|}{2 FC}                                                                & on-chip                                                           \\ \hline
\multicolumn{1}{|c|}{\cite{wang2019situ}}        & \multicolumn{1}{c|}{TaOx}                                                          & \multicolumn{1}{c|}{128x64}         & \multicolumn{1}{c|}{4b}                                                            & \multicolumn{1}{c|}{4b, 8b, 16b}                                                     & \multicolumn{1}{c|}{\begin{tabular}[c]{@{}c@{}}2 Conv, 1 FC\end{tabular}}             & on-chip                                                           \\ \hline
\multicolumn{1}{|c|}{\cite{xue2019embedded}}     & \multicolumn{1}{c|}{55nm}                                                          & \multicolumn{1}{c|}{156x512}        & \multicolumn{1}{c|}{1b}                                                            & \multicolumn{1}{c|}{3b, - (s-1-2b), 4b}                                              & \multicolumn{1}{c|}{-}                                                                   & PTQ                                                               \\ \hline
\multicolumn{1}{|c|}{\cite{yin2020high}}         & \multicolumn{1}{c|}{\begin{tabular}[c]{@{}c@{}}90nm, \\ HfO2\end{tabular}}         & \multicolumn{1}{c|}{128x64}         & \multicolumn{1}{c|}{1b}                                                            & \multicolumn{1}{c|}{1b, 1b$^{*3}$ , 3b}                                                & \multicolumn{1}{c|}{5 FC}                                                                & QAT                                                               \\ \hline
\multicolumn{1}{|c|}{\cite{hu2018memristor}}     & \multicolumn{1}{c|}{\begin{tabular}[c]{@{}c@{}}2um, \\ HfO2\end{tabular}}          & \multicolumn{1}{c|}{128x64}         & \multicolumn{1}{c|}{6b}                                                            & \multicolumn{1}{c|}{2b, 4b, 6b}                                                      & \multicolumn{1}{c|}{1 FC}                                                                & -                                                                 \\ \hline
\multicolumn{1}{|c|}{\cite{liu202033}}           & \multicolumn{1}{c|}{130nm}                                                         & \multicolumn{1}{c|}{784x100}        & \multicolumn{1}{c|}{1b}                                                            & \multicolumn{1}{c|}{1b(T), - (s-1b), 8b}                                             & \multicolumn{1}{c|}{3 FC}                                                                & -                                                    
\\ \hline

\multicolumn{7}{|l|}{\begin{tabular}[c]{@{}l@{}}s - serial input encoding 1b or 2b DAC; $^{*1}$W,I,O - weights, inputs, outputs;  \\ T - ternary; FC - fully connected layers; Conv - convolution layers;\\ $^{*2}$ several experiments; $^{*3}$ except input bits; $^{*4}$ pretrained with PTQ, 32b\end{tabular}}    
             \\ 
\hline

\end{tabular}
\label{t1}
\vspace{-0.7cm}
\end{table}

\vspace{-0.15cm}
\section{RRAM-based QNN hardware}
\vspace{-0.15cm}
%RRAM-based neural network architectures include RRAM crossbars for MAC operation and peripheral CMOS circuits including readout, control, and computation units. 
In the most common RRAM-based QNN implementations, only MAC operation is performed in the analog domain, while the other computations and control are digital. Such architectures include input encoding (DACs), analog 1T1R-based RRAM crossbar, multiplexers (MUX) and Trans-Impedance Amplifiers (TIA) for current to voltage conversion, Sample-and-Hold (S\&H), ADCs for converting MAC outputs to the digital domain, and Shift-and-Add (S\&A) operation \cite{chakraborty2020resistive}. 
In RRAM architectures, we focus on the quantization of network weights represented by $n$-bit RRAMs. Fig. \ref{f1} shows the summary of existing RRAM-based QNNs in terms of weight quantization relative to the device precision. It is also important to consider inputs quantization (DACs precision) and output quantization (ADCs precision). 

%MAYBE FIGURE! (Fig. \ref{f1} (c)). 

The calculation of partial sums in QNNs leads to a large ADC overhead. Partial sums are used for large weights matrices, bit slicing, and input slicing. 
QNN weight matrices larger than the number of rows in crossbar tiles are mapped to several RRAM cells to calculate partial sums followed by the addition of the output from several cells.
In bit slicing, high precision QNN weights are stored in several RRAM cells (usually different tiles), then partial sums from different tiles are shifted and added together.
%When QNN weight matrices are larger than the number of rows in crossbar tiles, they are mapped to several RRAM cells to calculate partial sums followed by adding the outputs from several cells.
%In bit slicing, high precision QNN weights are stored in several RRAM cells (usually different tiles), then partial sums from different tiles are shifted and added together. 
In input slicing, high precision inputs are streamed to a crossbar sequentially using low-bit DACs in several cycles (time encoding), and then summed together \cite{huang2021mixed}. The ADC resolution required to represent all possible MAC outputs depends on DAC resolution $b_a$, number of crossbar rows $i$ and bits in weights $b_w$, and can be expressed as $ceil(log_2((2^{b_a}-1)\times (2^{b_w}-1)\times i))$ \cite{chakraborty2020resistive}. 
ADCs consume up to 80-88\% of energy and 70-90\% of area of the crossbar with peripheral circuits \cite{ankit2019puma, yao2020fully}. 
To reduce ADC overhead, approximate computing with low-precision ADCs can be adopted or the number of ADCs can be reduced increasing the network latency \cite{zhang2020neuro}. The alternative approach reducing ADC overhead is post-MAC computations (e.g. activations) in the analog domain \cite{krestinskaya2020analogue}. However, the main issue in analog domain processing is a noise leading to error accumulation over entire network.

%FIGURE: In a crossbar with i rows for $b_a$-bit DAC and $b_w$-bit weights, ADC of $ceil(log_2((2^{b_a}-1)\times (2^{b_w}-1)\times i))$  bit resolution is required to represent all possible outputs (mixed-signal architectures) \cite{roy2020memory}. 

%Fig. \ref{f2} presents ADC resolution, which depends on DAC resolution $b_a$, number of crossbar rows $i$ and bits in weights  $b_w$, required to represent all possible MAC outputs.
%The ADC resolution required to represent all possible MAC outputs depends on DAC resolution $b_a$, number of crossbar rows $i$ and bits in weights $b_w$, can be be expressed as $ceil(log_2((2^{b_a}-1)\times (2^{b_w}-1)\times i))$ \cite{roy2020memory}.

%SHOULD BE ADDED?:
%For high-bit weights and inputs, e.g. 8-16 bits, the ADC resolution can still remain 8-16 bits without leading to accuracy degradation. While, low-bit architectures are more sensitive to ADC resolution degradation.

%HERE!!!

\begin{figure}[!t]
    %\centering
    \includegraphics[width=87mm]{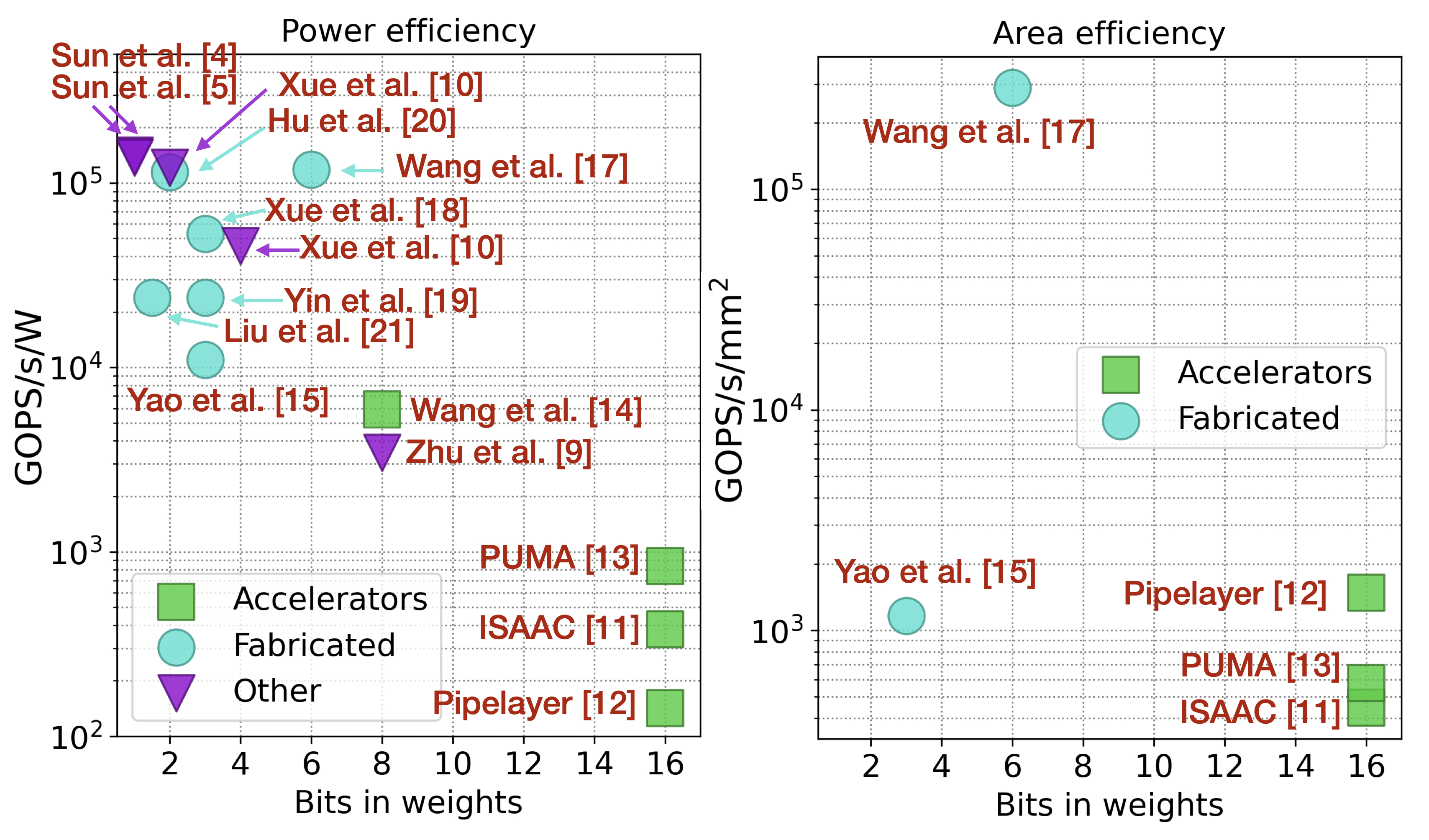}
    \caption{\footnotesize Power and area efficiency of state-of-the-art RRAM-based QNNs.}
    \label{f2}
    \vspace{-0.7cm}
\end{figure}

Table \ref{t1} shows the summary of the existing RRAM-based QNN architectures. Fig. \ref{f2} illustrates the power and area efficiency of these architectures.

\subsubsection{\textbf{Binary / Ternary RRAM Computation}}
In Binary networks (BNNs), the weights are presented by 1-bit devices corresponding to (+1,-1) software equivalent, and activations are also binarized. 
In BNNs, MAC operation is implemented via XNOR operation \cite{sun2018xnor}, and ADC overhead is reduced using a 1-bit sense amplifier (SA) \cite{sun2018fully}, as the activations are binarized. 
In \cite{9265062}, XNOR-based BNN is implemented on a 1T1R array with oxide semiconductor FET integrated into a 3D stack using a low-temperature BEOL process.
In \cite{yin2020high}, a fabricated BNN prototype (crossbar and peripheral readouts) is implemented with flash ADC, which is not practically scalable for higher bit precision networks.
In Ternary networks, the weights are set to (-1,0,+1) represented by two RRAM cells \cite{laborieux2020implementation}. 
In BNNs and Ternary networks training, the quantized weights and activations are used in the forward propagation, and quantized gradient values via STE in the backward pass. However, the full precision computation required for weight update and training complexity is the limitation for RRAM-based on-chip learning in BNNs.

%The input to the first BNN layer is non-binary.  
%However in practice, such implementation is prone to errors \cite{sun2018xnor}.
%BNN implementations are shown in \cite{sun2018xnor, yin2020high, 9265062,sun2018fully}.
%In \cite{sun2018fully}, binary CNN with parallel neurons readout and 1-bit SA instead of ADC is illustrated. However, the training of such architecture still requires 6-bit weights.

%In \cite{sun2018xnor}, XNOR-based BNN architecture is presented, where a 3-bit multi-level SA is used instead of noisy 1-bit SA.
%The example of XNOR-based Ternary RRAM network is shown in \cite{laborieux2020implementation}. 

%\cite{laborieux2020implementation} - 2 devices (-1,0,1), 
%can use precharge sense amplifier for reading weights, and ternary weight value can be extracted in a single weight operation, can be constructed using Gated XNOR operation (same as XNOR but with 0 possible outputs in MAC)

%--trained similar as binary: quantized weights in forwardpass and gradient values and quantized activations, STE in backward pass,  but full precision in update??  

%TERNARY:

%-- weight update is not ternary to keep good precision for the parameter update (gradients are identified using STE)

\subsubsection{\textbf{High Bits Fixed-point RRAM Computation }}
Compared to BNNs, the higher bit QNNs require ADCs and fixed-point computation in peripheral circuits \cite{kim20213,cai2019low,zhu2019configurable,kim20213}. 
In \cite{cai2019low}, the partial sums from large weight matrices split into crossbar tiles are quantized, and then merged via element-wise adder and quantized again aiming to reduce ADC overhead. These quantizations are also included in software-based training using floating-point weights and STE.

%The examples are shown in \cite{kim20213,cai2019low,zhu2019configurable}. In \cite{kim20213}, 3-bit QNN is shown. 

%\cite{cai2019low} - low bit CNN focusing on splitting large weight matrix to smaller crossbars, designing pipeline and including quantization to training (software). 

%In crossbar splitting to several tiles, the intermediate activation, is quantized to reduce ADC overhead. 
%The these intermediate activations are merged together using element-wise adder and quantized again. 
%In the training, the the splitting and intermediate activation quantization is included, while the backpropagation part is performed using floating-point operation and STE. The weights are quantized using power-of-two method. 

%\cite{kim20213} - illustrates Quantized MLP with 3-bit weights simulated based on the measurements from 32x32 fabricated $TiO_x/Al_2O_3$ crossbars. Using PTQ from 32b floating point weights. The conductance is distributed equally in the region between Ron and Roff.

\subsubsection{\textbf{Accelerator Level Design Considerations}}
In QNN accelerators based on fixed-point computation, the datapath, routing, and arrangement of network nodes are considered in addition to control and computation peripherals \cite{shafiee2016isaac,song2017pipelayer,ankit2019puma,ankit2020panther}. The inputs to most QNN accelerators are fed to the crossbar via low-precision DACs and serial encoding. In ISAAC accelerator \cite{shafiee2016isaac}, the weight partitioning and storing into several RRAM cells is considered. Also, weights are stored in either original or flipped form to increase the number of zero sums reducing the required ADC's precision. In Pipelayer architecture \cite{song2017pipelayer}, both training and inference are considered exploiting intra-layer parallelism. The PUMA architecture consists of nodes arranged to on-chip network including several multicores \cite{ankit2019puma}. A multicore consists of several cores and routing peripherals. While a core includes RRAM crossbar and peripherals, execution pipeline, and digital CMOS-based processing unit. The Panther architecture is based on PUMA but includes efficient training based on crossbar partitioning using the bit-slicing technique \cite{ankit2020panther}. Panther aims to update the crossbar with parallel writes encoding row signals in time-domain and column signals in amplitude domain via pulse-amplitude modulation.

%\cite{ankit2020panther} - Panther: RRAM based architecture with efficient training, partitioning crossbar using bit-slicing technique to ensure efficient training
%--aiming to realize the weights gradient and update without serial write
%--crossbar is updated encoding row signals in time-domain and column signal in amplitude domain using pulse-amplitude modulation during the update to ensure parallel update
%--the update power consumption grows with input resolution (DAC resolution) exponentially

%\cite{shafiee2016isaac} ISAAC - weight partitioning and storing into multiple 1T1R cells due to limited precision of the memristor
%--weights are stored in either original or flipped form to increase the number of zero sums, which can reduce the required precision of ADC

%\cite{song2017pipelayer} Pipelayer: supporting both training and inference, exploiting intra-layer parallelism 

%PUMA: consists of core (with RRAM crossbar and periferals, execution pipeline and digital CMOS-based processing unit), multicore (including several cores, shared memory and) , and routing (data movement) periferals), and node (including several nodes arranged to on-chip network). \cite{ankit2019puma}

\subsubsection{\textbf{Variable Precision and Dynamic Quantization}}
Variable (mixed) precision implies different quantization precision in different layers of QNN \cite{huang2021mixed}, as these layers have different quantization sensitivity \cite{zhu2019configurable}. Multi-precision QNNs are prone to hardware under-utilization and reconfigurability issues. In \cite{zhu2019configurable}, a multi-precision CNN accelerator with a different number of computing units of the same design used for variable precision is shown.
In dynamic quantization shown in \cite{chen2018high}, the number of analog-to-digital (AD) conversions and shift-and-add operations is reduced by skipping the computation of some partial products. Dynamic quantization is achieved by enabling a different number of columns in the crossbar for the AD conversion for different input bit positions.

\subsubsection{\textbf{Fabricated RRAM Crossbar-based Designs}}
By default, all fabricated RRAM architectures are quantized, as RRAM cells have limited precision. 
Recently, several QNN architectures with integrated RRAMs have been demonstrated for \cite{hu2018memristor} and tiles QNN architectures for inference \cite{wang2019deep,xue2019embedded}. The architecture in \cite{xue2019embedded} supports configurable input precision with bit-serial input encoding and shows a multi-bit processing trade-off between area and speed. In \cite{liu202033}, the readout ADC resolution can be adjusted by changing the sampling clock frequency.
In \cite{li2018efficient} and \cite{wang2019situ}, on-chip (in-situ) learning is considered, which requires additional hardware overhead for data routing and additional memory to store intermediate outputs. In-situ learning can compensate for RRAM  non-idealities and unresponsive devices. 
In \cite{li2018efficient}, RRAM crossbar is integrated to customized PCB, while computation of activations and backpropagation is performed in software. In \cite{wang2019situ}, the in-situ training of CNN and ConvLSTM is shown focusing on shared weights-based architectures. In \cite{yao2020fully}, the hybrid training to accommodate device variations is proposed. The network is trained off-chip followed by on-chip fine-tuning, where a single fully-connected layer is updated to avoid complex backpropagation, which compensates for errors caused by variations in duplicated convolution kernels.

\begin{table}[t]
%\centering
\scriptsize \setlength{\tabcolsep}{0.1pt}
\renewcommand{\arraystretch}{0.75}
\caption{QNNs: Open Challenges and Requirements}
\begin{tabular}{|ll|}
\hline
\multicolumn{1}{|c|}{Device level requirements}                                                                                                                                                          & \multicolumn{1}{c|}{Architecture and Algorithmic Support Requirements}                                                                                                                                                                                                                                                                                                                                    \\ \hline
\multicolumn{2}{|l|}{\cellcolor{green!10} \textbf{ Open Challenge: Efficient Inference}}                                                                                                                                                                                                                                                                                                                                                                                                                                                                                                                                                                     \\ \hline
\multicolumn{1}{|l|}{\begin{tabular}[c]{@{}l@{}}Reducing: \\ $\bullet$ C2C and D2D variations\\ $\bullet$ Conductance drift with time\\ $\bullet$ Device faults\end{tabular}}                                                             & \begin{tabular}[c]{@{}l@{}} $\bullet$ Reducing ADC and partials sums overhead\\ $\bullet$ Improving data movement between the layers\end{tabular}                                                                                                                                                                                                                                                                                   \\ \hline
\multicolumn{2}{|l|}{\textbf{\cellcolor{green!10} Open Challenge: Training/Learning On-Chip}}                                                                                                                                                                                                                                                                                                                                                                                                                                                                                                                                                               \\ \hline
\multicolumn{1}{|l|}{\begin{tabular}[c]{@{}l@{}} $\bullet$ Improving device \\ endurance and faults\\ $\bullet$ Improving linearity and \\ symmetry of conductance \\ updates\\ $\bullet$ Reducing high write cost\end{tabular}}           & \begin{tabular}[c]{@{}l@{}} $\bullet$ Developing efficient additional (higher) precision\\  support for training\\ $\bullet$ Developing new training algorithms to reduce \\ quantization precision\\ $\bullet$ Reducing CMOS and periferals overhead in training\\ $\bullet$ Solving large shared memory requirement issues\\ $\bullet$ Overcoming training latency via parallel crossbar \\writes\end{tabular}                                              \\ \hline
\multicolumn{2}{|l|}{\cellcolor{green!10} \textbf{Open Challenge: Reconfigurability, Towards General-Purpose Architectures}}                                                                                                                                                                                                              \\ \hline
\multicolumn{1}{|l|}{\begin{tabular}[c]{@{}l@{}} $\bullet$ Fulfilling RRAM devices\\  requirements with higher \\ bit precision and low \\ variations (as one solution\\  for variable precision \\ support)\end{tabular}} & \begin{tabular}[c]{@{}l@{}} $\bullet$ Developing different workloads support\\ $\bullet$ Developing variable precision support and reducing \\ corresponding hardware overhead\\ $\bullet$ Improving hardware utilization (balancing execution \\ time and energy consumption)\\ $\bullet$ Developing reconfigurability schemes for multi-bit \\ precision without hardware overhead\\ $\bullet$ Developing flexible control and data movement\end{tabular} \\ \hline
\multicolumn{2}{|l|}{\cellcolor{green!10} \textbf{Open Challenge: RRAM-based Solutions Integration and Practical issues}}                                                                                                                                                                  \\ \hline
\multicolumn{1}{|l|}{\begin{tabular}[c]{@{}l@{}} $\bullet$ Developing large scale\\  implementation\\ $\bullet$ Improving 3D integration\end{tabular}}                                                                             & \begin{tabular}[c]{@{}l@{}} $\bullet$ Testing on large databases + software-hardware \\ co-design\\ $\bullet$ Integrating RRAM-based solutions into traditional \\ processing architectures\end{tabular}                                                                                                                                                                                                                               \\ \hline
\end{tabular}
\label{t2}
\vspace{-0.7cm}
\end{table}

\vspace{-0.2cm}
\section{Open challenges and requirements}
\vspace{-0.2cm}
While most of the implemented QNN architectures focus on inference optimization, the problems, such as on-chip training, reconfigurability supporting different precision and workloads aiming for general-purpose chips \cite{zhang2020neuro}, and large scale implementation, remain open.
Table \ref{t2} summarizes the open research directions and corresponding challenges and requirements related to device and architecture level limitations.

\subsubsection{\textbf{Efficient inference}}
Regardless of the number of works on RRAM-based QNN inference, the inference efficiency can still be improved by reducing the number of ADCs, lowering ADC precision relying on approximate computing, or designing low-power application-specific converters. 
The larger crossbar tiles can reduce ADC overhead \cite{yao2020fully}, however longer wires contribute to increased IR drop \cite{zhu2019configurable}. 
This can be solved by developing 3D integration techniques and improving lithography techniques for 3D stacking \cite{chakraborty2020resistive}, which is also affected by selector device switching and non-linearity.
Partitioning large weights matrices or high-bit precision computations to partial sums also contributes to storage and datapath overhead \cite{chakraborty2020resistive}.
D2D variations may lead to errors accumulated and propagated over the network.
The device conductance variability depends on the selection of the RRAM material stack and can be addressed by on-chip fine-tuning.

%\cite{kim20213} - possible to reduce wire resistance by engineering the electrode metal stack

%\cite{li2018review} : $Ron/Roff=10-10^3$, $Ron=10^3-10^4$, $tset/reset$ can reach as low as 50ns, and the endurance reaches up to $10^6$ cycles, and retention time up to $10^8$s.

\subsubsection{\textbf{On-chip training}}
From the device perspective, efficient in-situ on-chip training is an open challenge due to limited device endurance and expensive crossbar writes. 
For example, training of ResNet-50 for ImageNet classification, requires 18M crossbar writes \cite{chakraborty2020resistive}. For $50ns$-writes, it would take $0.8s$ to update the whole network sequentially. Thus, parallel crossbar writes are essential for a large network, while they would contribute to hardware overhead.
The endurance of currently available devices reaches up to $10^6$ cycles, while at least $10^8-10^9$ cycles are required for on-chip training.
Non-linear asymmetric conductance update can also affect the on-chip learning schemes and would require additional overhead to overcome it, e.g. via read-verify-write mechanism \cite{chakraborty2020resistive}.

From the architecture perspective, the hardware overhead related to on-chip learning includes external memory for intermediate outputs storage, modules for additional computation requiring high precision, and flexible crossbar peripherals. Even for hybrid training (fine-tuning), the peripherals overhead is unavoidable due to batch training and the requirement to store intermediate results \cite{yao2020fully}. Moreover, 
most of the QNNs in Table \ref{t1} still require 32-bit weights and 
computation precision to train the networks, while low-bit precision is sufficient for the inference. This leads to the development of flexible precision support by either adjusting the number of RRAM cells representing a single weight or increasing the number of stable conductance levels in RRAM devices. Both contribute to increasing bit-precision requirements for ADCs. The alternative is to develop QNN training algorithms based on low-bit weights and low-bit arithmetic, which would lead to a significant reduction in hardware overhead for the training.

%Flexible precision can be achieved by either adjusting the number of RRAM cells representing a single weight, or increasing number of stable conductance levels in RRAM devices, which would in turn lead to increase ADC precision requirements.
%The other solution is to develop QNN training algorithms based on low-bit weights and low-bit arithmetic.

%\cite{yao2020fully} -- to increase the accuracy, higher bit precision of weights is required leading to increased hardware cost
%--Even for tuning, still additional memory blocks and data-transmission modules are required due to batches and requirement to store intermediate results

\subsubsection{\textbf{Reconfigurability and general-purpose architectures}}
Most RRAM-based QNN architectures are application-specific and designed for certain QNN types.
Reconfigurability is a major step towards general-purpose architectures supporting different workloads, multi-bit precision, flexible control and data movement, and efficient hardware utilization. 
Depending on the type of the executed layers in QNN or types of networks, the workloads vary. This leads to an unbalanced execution time in different layers \cite{zhang2020neuro} and the requirement to design flexible interconnects. For example, CNN workloads may be very different from sparse fully-connected layers or temporal processing in LSTM. The generalized reconfigurable accelerator should be able to support the processing in both spatial and temporal domains. Therefore, to achieve continuous data flow, balancing the execution time of different layers is required.
The other challenge arising from different workloads is a resource utilization balance. Mapping the network to a general-purpose RRAM accelerator can lead to resource under-utilization \cite{qu2020raqu}.
The reconfigurability of the architecture is also related to flexible bit-precision support in different QNN layers and minimizing its overhead. Also, the hardware-aware algorithms automating required precision computation based on the accuracy and energy trade-offs in different layers can be further developed.

%Flexibility and reconfigurability of the architecture can be determined by a support of different workloads, variable multi-bit precision in different layers, flexible control and data movement, and efficient hardware utilization. 

%The reconfigurability of the architecture is also related to flexible bit-precision support in different QNN layers. 

%Flexible bit-precision of RRAM cells is required for multi-bit layers implementation, and higher precision training. 
%Flexible arbitrary precision can be achieved by either adjusting the number of RRAM cells representing a single weight, or increasing bit-precision of RRAM by improving material stack \cite{qu2020raqu}.

%In RRAM-based QNNs, the trade-off between data reuse, energy efficiency and tolerance to variations should be considered.

\subsubsection{\textbf{Efficient integration of RRAM-based solutions into the traditional processing hierarchies}}

Considering the RRAM limitations, the complexity of QNN training and the demand for moving toward general-purpose architectures, the efficient integration of RRAM-based solutions with traditional memory and processing hierarchies is essential. For example, for the on-chip training implementation, the efficient integration with traditional memories is important to overcome external memory requirements and latency issues. The example of such a system is shown in \cite{chang202240nm}, where RRAM-based processing is integrated with conventional SRAM-based memory for AI applications. Overall, large-scale integration is the next step to consider in the RRAM-based AI hardware development.

\vspace{-0.15cm}
\section{Conclusion}
\vspace{-0.15cm}
This paper reviews state-of-the-art RRAM-based QNN architectures and corresponding device and architecture level metrics. We emphasize possible future research directions and open challenges related to the improvements of QNN inference efficiency, on-chip training, architecture reconfigurability, and demands for the efficient integration of RRAM-based solutions into traditional processing hierarchies.

\vspace{-0.15cm}
\setstretch{0.7}
%\balance
\bibliographystyle{IEEEtran}
\bibliography{ref}

\end{document}